**Screenline-based Two-step Calibration and its application to an agent-based urban freight simulator**

Yusuke Hara[a]*, Takanori Sakai[b], André Romano Alho[c], Moshe Ben-Akiva[c]

## ABSTRACT

Calibration is an essential process to make an agent-based simulator operational. Especially, the calibration for freight demand is challenging due to the model complexity and the shortage of available freight demand data compared with passenger data. This paper proposes a novel calibration method that relies solely on screenline counts, named Screenline-based Two-step Calibration (SLTC). SLTC consists of two parts: (1) tour-based demand adjustment and (2) model parameter updates. The former generates screenline-based tours by cloning/removing instances of the simulated goods vehicle tours, aiming to minimize the gaps between the observed and the simulated screenline counts. The latter updates the parameters of the commodity flow model which generates inputs to simulate goods vehicle tours. To demonstrate the practicality of the proposed method, we apply it to an agent-based urban freight simulator, SimMobility Freight. The result shows that SLTC allows the simulator to replicate the observed screenline counts with reasonable computational cost for calibration.

[a] Graduate School of Information Sciences, Tohoku University

6-6-06, Aramaki-Aoba, Aoba-ku, Sendai, Miyagi, Japan.

[b] Department of Logistics and Information Engineering, Tokyo University of Marine Science and Technology, Japan

2-1-6, Etchujima, Koto-ku, Tokyo, Japan.

[c] Intelligent Transportation Systems Lab, Department of Civil and Environmental Engineering, Massachusetts Institute of Technology

77 Massachusetts Avenue, Room 1-181, Cambridge, M.A., 02139, United States

Email address

| | |
|---|---|
| Yusuke Hara: | hara@tohoku.ac.jp |
| Takanori Sakai: | tsakai2@kaiyodai.ac.jp |
| André Romano Alho: | aromanoa@mit.edu |
| Moshe Ben-Akiva: | mba@mit.edu |

* Corresponding author.

| | |
|---|---|
| Postal Addresses: | 6-6-06, Aramaki-Aoba, Aoba-ku, Sendai, Miyagi, Japan. |
| E-mail Address: | hara@tohoku.ac.jp |
| Telephone Number: | +81-22-795-4419 |

# 1. Introduction

In recent years, there has been remarkable progress in the capabilities of transportation simulators, leading to an increasing role for disaggregate, agent-based models (ABMs). For simulating activity-based passenger movements using private vehicles and public transport, a number of ABMs were proposed including MATSim (Horni et al., 2016), CEMDAP (Pinjari et al., 2008), POLARIS (Auld et al., 2016), SimMobility (Adnan et al., 2016), and CityMoS (Zehe et al., 2017). Similar developments also occurred for urban freight. ABMs for urban freight have been proposed to replace traditional aggregate commodity- or truck-based models, replicating the behavior of agents across supply chains and associated logistics operations (Chow et al., 2010). In these models, varying types of agents, such as suppliers, receivers, carriers, drivers and policy makers, are considered in the process to predict commodity flows, logistics/transportation services, vehicle flows, and transportation infrastructure usage (Wisetjindawat et al., 2005; Fischer et al., 2005; Roorda et al., 2010; de Bok and Tavasszy, 2018; Sakai et al., 2020).

ABMs are usable for testing a wide range policies and technologies. Aiming to produce realistic cases, the calibration of the behavioral model parameters and associated travel demand is required. However, the complexity associated with increased realism of ABMs escalates the difficulty in calibrating model parameters. Firstly, the execution time of a simulation is comparatively longer in a disaggregate model compared to an aggregate model. Secondly, these models require various types of data with large samples to properly adjust the parameters for each model component. However, the collection of disaggregate data regarding the decisions of agents is a challenging and resource-consuming task. Therefore, the calibration often needs to rely on aggregate data. Lastly, the high degrees of freedom associated with the parameters in an ABM makes the computational cost extremely high to optimize them using existing methods. These difficulties become critical to transfer an ABM from one city to another, which would require a new set of data and a new calibration process. These challenges prompt for the need of an efficient calibration method.

We propose a novel calibration framework, named Screenline-based Two-step Calibration (SLTC), targeted at transportation-focused ABMs dealing with tour-based travel demand and its assignment to transportation network using traffic simulation. We define "tour" as a sequence of trips and, thus, as a sequence of nodes. Unless otherwise noted, a "tour" in this paper indicates a sequence of nodes without the information of routes. The proposed framework aims to achieve low computational cost and be implementable with the minimum data requirement, i.e., screenline count data. Screenline count data is defined as traffic counts along lines crossing one or more road sections. The application of SLTC generates quasi-observed tours, called *Target Tours*, based on the simulated tours and screenline count data. *Target Tours* are, in turn, used to update parameters of demand models (which can be regressions, discrete choice models, among other options). To demonstrate the practicality of the framework, we apply SLTC to an urban freight model, SimMobility Freight (Sakai et al., 2020).

The rest of the paper is organized as follows: Section 2 discusses existing calibration and demand synthesis methods; Section 3 describes the details of SLTC; Section 4 presents a demonstrative application of SLTC to an agent-based urban freight model; and Section 5 concludes the research with remarks about the contribution and the future research.

# 2. Existing calibration and demand synthesis methods

2.1. Calibration methods

According to Rakha et al. (1996) and Hellinga (1998), model calibration is considered as the process of determining to what extent the model user can, or is required to, modify the default input parameter values, that describe the underlying mechanics, in order to reflect the observed local traffic conditions being modeled. Common practice is to change the original model parameters, as little as possible, based on empirical data for matching observed and simulated traffic. In general, there are four main factors that make the calibration challenging. First, available data are often limited in terms of both quality and quantity. For instance, while sensor and link travel time data are the most common data available for transport systems in an urban environment, these are usually only available for limited sections in transportation network. Further, such data is not useful to infer trip Origins/Destinations (OD) nor to determine factors on the choices of agents (e.g., driver's routing choices). Second, an ABM for transportation simulations is a non-linear system, in which the optimum solutions are difficult to be identified. Third, the calibration problem of an ABM can entail high computational cost due to the model complexity and associated run times which span across hours and/or days. Lastly, particularly for freight models, the difficulty in business-level data collection often results in a smaller number of available measurements compared to those available for a passenger model.

There are several studies on the methods for calibration of OD-level travel demand, the most conventional inputs for traffic simulations. The process of OD estimation/calibration aims at a high fit between the observed and simulated traffic flows. However, the problem is often ill-posed because the number of OD pairs is much greater than that of screenlines for traffic counts. Therefore, a number of studies focus on removing the bias of the ill-posed problem. Cascetta et al. (1993) propose a method to estimate or update OD demand based on traffic counts. They use an assignment fraction matrix, which captures the relationship between OD demand and link flows, both of which are dynamic by time-of-day. Ashok and Ben-Akiva (2000) propose the real-time estimation of time-dependent OD demand using state-space models. The same authors also propose a method of the stochastic mapping from OD demands to link flows (Ashok and Ben-Akiva, 2002).

The calibration of model parameters leading to the ODs is also an important research subject. For traffic simulators, the Simultaneous Perturbation Stochastic Approximation (SPSA) algorithm (Spall, 1998) is one of most promising approaches. Weighted Simultaneous Perturbation Stochastic Approximation (W-SPSA) (Antoniou et al., 2015) is an extension of SPSA which aims at achieving greater efficiency. Since some variables are strongly correlated in models, the determination of a proper weight matrix improves the calibration process. However, both approaches require an execution of the simulator at each iteration of the algorithm and for this reason the approaches are computationally inefficient and thus impractical. In this paper we propose a calibration method which is capable of addressing cases where: (1) the combination of OD pairs is much greater than the number of screenlines for traffic counts, and (2) a large number of simulation iteration runs is impractical.

2.2. Freight demand synthesis

The research on freight demand (origin-destination and tour) synthesis is worth mentioning here as the proposed SLTC can be used as an approach for the demand synthesis. Origin-destination synthesis (ODS), an approach to generate origin-destination matrices based on traffic count data, has been widely used and extended mainly for passenger trips, although the research which considers the characteristics of freight demand is limited. Holguín-Veras and Patil (2007) proposes a goods vehicle ODS method based on a gravity model considering commodity flows (with a single commodity type), load factors, and empty trips. This method uses traffic counts and freight generation (production and consumption) as inputs. Holguín-Veras and Patil (2008) expands the model formation to consider multiple commodity types and demonstrates that the multi-commodity model outperforms the single-commodity one. Recently, Malik et al. (2021) proposes a framework for urban freight ODS with limited data. The authors use traffic count as primary data while use real-time link speeds obtained from Google's API as secondary data. The framework involves multiple steps including the calibration of speed-volume relationships, the estimation of direct flows to complement the count data before applying Nielson's multimodal ODS model equipped in TransCAD.

On the other hand, ODS method is considered problematic as it ignores trip sequences. In the real world the significant share of urban goods vehicle trips is associated with multiple stop delivery/pickup tours (Wang and Holguín-Veras, 2009; Sánchez-Díaz et al., 2015; Gonzalez-Calderon and Holguín-Veras, 2019) that motivates the research to develop freight tour synthesis (FTS). Wang and Holguín-Veras (2009) first propose entropy maximization formulations to estimate tours based on trip generations at nodes and the total impedance of the network. Sánchez-Díaz et al. (2015) extend their approach with the additional use of traffic count data and propose a time-dependent freight tour synthesis (TD-FTS) model, acknowledging the limitation of the classic ODS, which do not consider tours, for the purpose of urban freight analysis. Their method uses as inputs, traffic counts per time interval, (goods vehicle) trip generation, and the total freight transportation cost in the network, to generate synthetic freight tours. They focus on goods vehicle tours and trip generation needs to be estimated exogenously (thus, the trip generation model needs to be reasonably calibrated in advance.). Gonzalez-Calderon and Holguín-Veras (2019) formulate a non-time-dependent FTS and use it to evaluate multiple heuristics to determine traffic count locations in the experiment using Sioux Falls network.

The above existing FTS methods do not consider the relationship between commodity flows and vehicle trips and should rely on the accuracy of goods vehicle trip generations. On the other hand, the approach described later in paper, the use SLTC on SimMobility Freight, produces both commodity flows and goods vehicle tours, while calibrates the model parameters for commodity flows, using traffic count data and prior model parameters.

### 3. Method

3.1. Problem setting

The objective of the SLTC method is to calibrate tour-based demand models used in an ABM, as to minimize the gaps between observed and simulated screenline counts. Examples of tour-based travel demand are day activity patterns of individuals (Bowman and Ben-Akiva, 2001) and tour-based goods vehicle movements (Hunt and Stefan, 2007; de Bok and Tavasszy, 2018; Sakai et al., 2020). The method is suitable for a context where an ABM first outputs predictions of tour-based travel demand, which are then assigned to the network. Once assigned to the network, agents perform route choices which, when aggregated, can be compared with screenline counts. We assume a situation where an ABM has initial model parameters estimated based on empirical data and observed screenline count data is available.

Figure 1 provides an overview of SLTC process. The calibration problem is divided into two sub problems: (1) tour-based demand adjustment and (2) model parameter updates. The process of tour-based demand adjustment generates *Target Tours* by cloning/removing the simulated tours with the objective of minimizing the gaps between observed and simulated screenline counts. In this context, and in general terms, cloning means duplicating a given tour record, and removing means deleting the tour record from the estimated demand. Additional details on the cloning process will be further provided. The output of this process - *Target Tours* - are used as a substitute for empirical data to update model parameters. The process of updating model parameters is achieved by re-estimating model parameters using "quasi-observed" data which generated based on *Target Tours* and can rely on various levels of aggregation. It should be noted that the actual formulation of this sub-problem depends on the structure of the demand models that are present in a given ABM. In the proposed framework, the two sub problems are repeatedly solved in sequence until the gap between the observed and simulated screenline counts reaches a level of acceptance which does not require further demand adjustment.

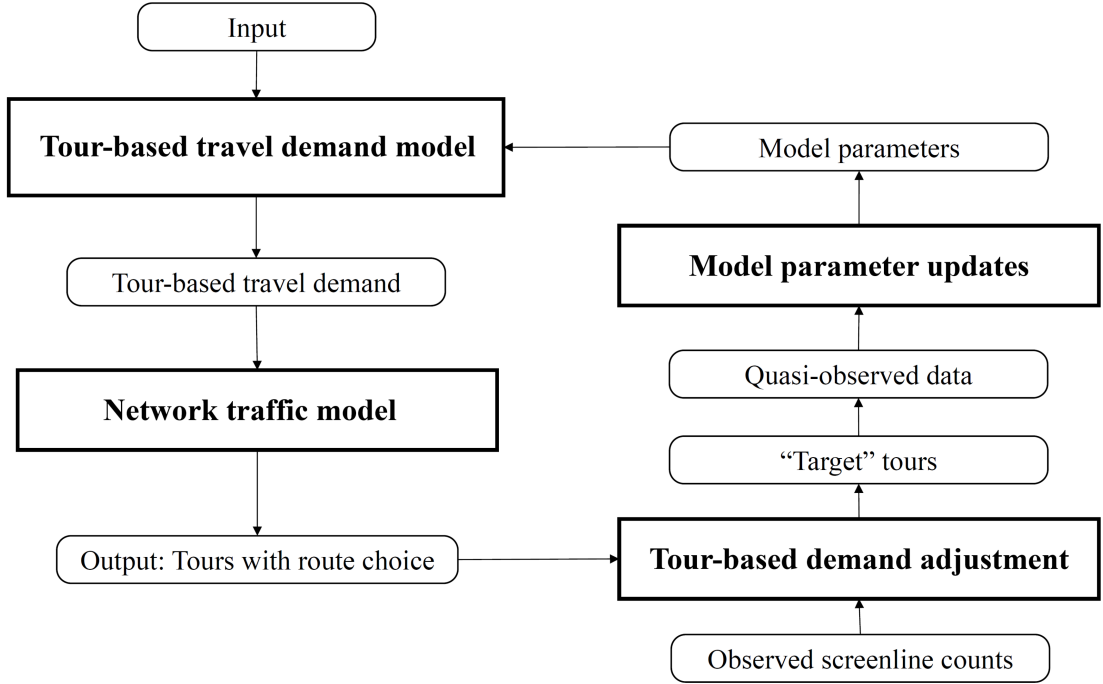

**Fig. 1** The overview of Screenline-based Two Step Calibration. The left side is the flow of a simulation model and the right side is the flow of proposed calibration method.

3.2. Tour-based demand adjustment

*3.2.1. Screenline-based tours*

We first define a novel concept of a tour, denominated as the screenline-based (SLB) tour. A SLB tour is defined by a sequence of (crossed) screenlines, instead of a sequence of nodes or trips. By design, the generation of SLB tour requires a node-based tour to be assigned to the transportation network and route choice simulations. A pair of non-identical node-based tours could be identical SLB tours in some cases. Figure 2 illustrates this concept. There are five tours in the figure. Two of these tours pass through screenlines A and B, labeled as SLB tour a. Other two tours pass through screenlines B and D, labeled as SLB tour b. Lastly, the tour passes through screenlines C and D, labeled as SLB tour c. In summary, there are three unique SLB tours – SLB tours a, b, and c - and their counts are 2, 2, and 1, respectively.

The relationship between SLB tours and simulated screenline counts is as follows. Let $K$ denote the set of screenlines. Each screenline $k \in K$ has a screenline count $y_k^o$. The observed screenline count vector is $y^o = (\dots, y_k^o, \dots)$. Given all tour routes, which are available by running a route choice model, we can enumerate all unique SLB tours. Let $L$ denote the set of unique SLB tours. Each unique SLB tour $l$ has its count $x_l^o$ and the vector of these counts is $x^o = (\dots, x_l^o, \dots)$. Also, we can define the mapping matrix for the relationship between unique SLB tours and screenlines; if a SLB tour passes through a screenline, the value of the corresponding element is 1; 0 otherwise. The mapping matrix is $A \in R^{|L| \times |K|}$ and $A^T x^o$ indicates the vector of simulated screenline counts. The right side of Figure 2 shows the relationship between screenline counts, mapping matrix, and SLB tour counts. Though the number of unique SLB tours $|L|$ is only 3 in this example, $|L|$ is generally large when applying this concept to a city-scale simulation.

A major advantage of using the concept of SLB tour is that the number of unique SLB tours is much smaller than that of OD pairs or node-based tours in a typical ABM, which will be further illustrated in the demonstration later in this paper. Since the efficiency of calibration method depends on the number of manipulated variables, the use of SLB tours is comparatively more efficient. As SLB tours are to be generated by using a route choice model, the validity of the relationship between node-based tours and SLB tours can be assured by the validity of such model. If we assume that the vehicles in same (node-based) tours tend to take same routes at a given time, the re-run of a routing model is not necessary; this assumption can significantly reduce the computational cost required in the calibration procedure.

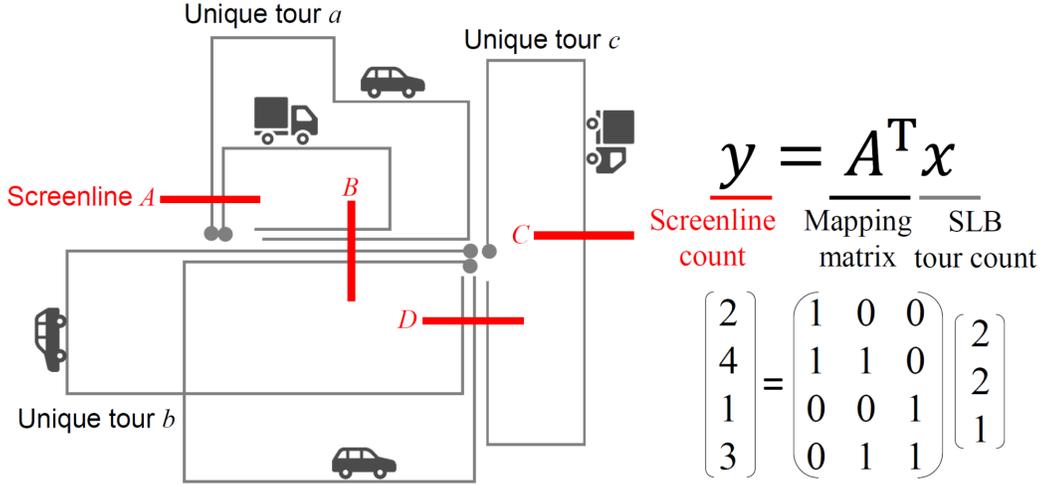

**Fig. 2** An illustrative example of screenline-based tour concept.
The left side shows the five vehicle's tours and four screenlines. The right side shows the relationship between screenline counts and unique SLB tour counts.

*3.2.2. Problem definition*

In order to obtain *Target Tours*, we assume that the routes of tours are the same after the adjustment (i.e. cloning and removing tours) as mentioned earlier, simplifying the problem for efficient computation. The objective of the tour-based demand adjustment is to minimize the gap between the observed and simulated screenline counts by cloning or removing tours in the tour-based travel demand, using the concept of SLB tour. We employ a square error as the error indicator and $x$ indicates the change in the numbers of SLB tours to be made (i.e. SLB tours are cloned if positive and removed if negative).

The problem is defined as below:

$$\begin{aligned} x^* &= \arg\min_{x} \lVert y^o - A^T x^0 - A^T x \rVert^2 \\ &= \arg\min_{x} \lVert y - A^T x \rVert^2, \end{aligned} \qquad (1)$$

where $y \equiv y^o - A^T x^0$ which indicates the gaps between the observed and simulated screenline counts. To solve this problem easily, we employ the continuous relaxation of $x$.

The solution of the problem is:
$$x^* = (AA^T)^{-1} Ay. \qquad (2)$$

The solution of the above equation could require significant changes (and over-fitting) in the tour-based demand, which would lead to a perfect match between the post-adjustment simulated screenline counts and the observed screenline counts. To avoid such situation, the constraint is required to keep x reasonably small.

$$x^* = \arg\min_{x} ||y - A^T x||^2 + \lambda ||x||^2, \tag{3}$$

where $\lambda$ is a penalty parameter, which can be interpreted as L2 regularization. This problem is easy to solve as this equation is equivalent to Ridge regression (Hoerl and Kennard, 1970), which is the regression model with L2 regularization.

The objective function is defined as follows:

$$\begin{aligned} J &= (A^T x - y)^T (A^T x - y) + \lambda x^T x \\ &= (x^T A - y^T)(A^T x - y) + \lambda x^T x \\ &= x^T A A^T x - 2 x^T A y + y^T y + \lambda x^T x. \end{aligned} \tag{4}$$

To minimize the objective function, the solution must satisfy the first order condition.

$$\frac{\partial J}{\partial x} = 2 A A^T x - 2 A y + 2 \lambda x = 0. \tag{5}$$

The solution of the problem is derived as:

$$x^* = (A A^T + \lambda I)^{-1} A y. \tag{6}$$

The important point is that $AA^T$ is a positive semidefinite matrix and $\lambda I$ is a positive definite matrix when $\lambda > 0$; therefore, the matrix $(AA^T + \lambda I)^{-1}$ always exists. The solution $x^*$ is presented in continuous values, and thus not directly the numbers of SLB tours to clone or remove. To obtain integer values, the solution must be rounded. After rounding, a positive value indicates the number of SLB tours to clone and a negative value those to remove. If the number of SLB tour i to remove is greater than its simulated number, the process is redone with the additional condition of fixing $x_i$ as $(-1) \times$ (the simulated number of SLB tour i).

Finally, the solution $x^*$ is used to adjust the tour-based demand. For cloning or removing tours for a specific SLB tour, the original (i.e. node-based) tours which are associated with the SLB tour of interest are randomly selected. The output of this process, i.e. the adjusted tour-based demand, is denominated as *Target Tours*.

While a discussion of the temporal dimension is omitted for simplification, the above tour-based demand adjustment can be applied for the different levels of temporal granularity (e.g. hourly, AM/PM) although in this case it is demonstrated in an application to daily traffic counts.

*3.2.3. Model parameter updates*

*Target Tours* can be used, in turn, as quasi-observed data for updating model parameters to predict tour-based demand. *Target Tours* consists in a set of tours associated with an individual/vehicle as well as departure time from an origin, the arrival time at a destination, trip purpose, transport mode, and other attributes of the trips. *Target Tours* can be used for (re-)estimation of parameters in disaggregate models, including discrete choice models such as destination choice and departure time choice models. Contrasting with aggregated distribution data (i.e. OD table) which is often not useful for parameter estimation of a destination choice model for an ABM, the *Target Tours* include destination choices as well as the information of each individual/vehicle and thus allow for updating parameters for disaggregate models.

*3.2.4. Theoretical discussion on assumptions*

In this subsection, we discuss the assumptions associated with SLTC. As a transportation simulation with an ABM is a direct problem, we can describe a probabilistic graphical model of the simulation process. Figure 3 expresses the conditional dependence structure between random variables. In Figure 3a, Model 1 (parameter $\theta_1$) generates intermediate variables $Z_1$, Model 2 (parameter $\theta_2$ and $Z_1$) generates $Z_2$, and Model 3 (parameter $\theta_3$ and $Z_2$) generates final output X. In our setting, X is screenline counts. Following the typical graphical model representation, white circles indicate unobservable variables and gray circles indicate observable variables.

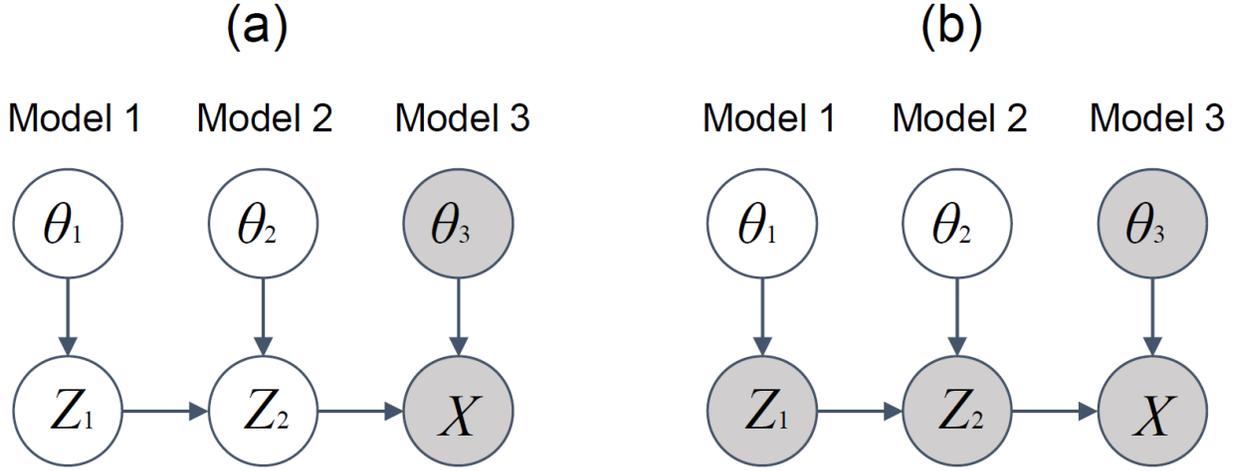

**Fig. 3** The graphical representation of a simulation model.
The left figure (a) indicates the case that we observe final output $X$ only. The right figure (b) indicates the case that we additionally observe intermediate variables $Z_1$ and $Z_2$.

We are interested in the parameter estimation process using observed data $X_{obs}$. For simplicity, we assume we know the model parameter $\theta_3$ of Model 3. In the context of a transportation model, this situation is such that we know the model parameters of the network traffic model. Given $X_{obs}$, $\widehat{\theta_3}$ and the dependency defined above, the posterior joint probability distribution is as follows:

$$P(\theta_1, \theta_2, Z_1, Z_2 | X, \theta_3) = P(\theta_1 | Z_1, Z_2, X, \theta_3) \cdot P(Z_1 | Z_2, X, \theta_3) \cdot P(\theta_2 | Z_2, X, \theta_3) \cdot P(Z_2 | X, \theta_3), \quad (7)$$

where $P(\cdot | X, \theta_3) \equiv P(\cdot | X = X_{obs}, \theta_3 = \widehat{\theta_3})$ for simplicity. The equation is derived from the chain rule of probabilities. Therefore, the posterior probability distributions of model parameters $\theta_1$ and $\theta_2$ are:

$$P(\theta_1 | X, \theta_3) = \int_{Z=(Z_1, Z_2)} \int_{\theta_2} P(\theta_1, \theta_2, Z_1, Z_2 | X, \theta_3) \, d\theta_2 dZ$$
$$= \int_Z \int_{\theta_2} P(\theta_1 | Z_1, Z_2, X, \theta_3) \cdot P(Z_1 | Z_2, X, \theta_3) \cdot P(\theta_2 | Z_2, X, \theta_3) \cdot P(Z_2 | X, \theta_3) \, d\theta_2 dZ, \quad (8)$$

$$P(\theta_2 | X, \theta_3) = \int_{Z=(Z_1, Z_2)} \int_{\theta_1} P(\theta_1, \theta_2, Z_1, Z_2 | X, \theta_3) \, d\theta_1 dZ$$
$$= \int_Z \int_{\theta_1} P(\theta_1 | Z_1, Z_2, X, \theta_3) \cdot P(Z_1 | Z_2, X, \theta_3) \cdot P(\theta_2 | Z_2, X, \theta_3) \cdot P(Z_2 | X, \theta_3) \, d\theta_1 dZ. \quad (9)$$

From the perspective of Bayesian estimation, the objective is to obtain the posterior model parameter distribution itself: $P(\theta_1 | X, \theta_3)$ and $P(\theta_2 | X, \theta_3)$. From the perspective of maximum likelihood estimation, the objective is to obtain the model parameters $\widehat{\theta_1}$ and $\widehat{\theta_2}$ to maximize the likelihood.

$$\widehat{\theta_1} = \arg\max_{\theta_1} P(\theta_1|X,\theta_3) = \int_Z \int_{\theta_2} P(\theta_1,\theta_2,Z_1,Z_2|X,\theta_3)\ d\theta_2 dZ, \qquad (10)$$

$$\widehat{\theta_2} = \arg\max_{\theta_2} P(\theta_2|X,\theta_3) = \int_Z \int_{\theta_1} P(\theta_1,\theta_2,Z_1,Z_2|X,\theta_3)\ d\theta_1 dZ. \qquad (11)$$

In any case, it is difficult to obtain the parameter distributions or estimate the parameters because the solution space of each parameter is large, and we need to calculate the integral despite having no efficient calculation method. Similarly, it is difficult to estimate other intermediate variables $Z_1$ and $Z_2$.

The approach of SLTC, generating quasi-observations of $Z_1$ and $Z_2$ from *Target Tours* to substitute them for observations, mimics the situation in Fig 3(b). While this substitution (or approximation) entails a strong assumption, it is reasonable in practice if the original model parameters are estimated using suitable empirical data.

Given $X_{obs}, \theta_3$ and quasi-observations $z_1^*$ and $z_2^*$, the probability distributions of posterior parameter are as follows:

$$P(\theta_1|Z_1,Z_2,X,\theta_3) \approx \int_{\theta_2} P(\theta_1|Z_1,Z_2,X,\theta_3) \cdot P(\theta_2|Z_2,X,\theta_3)\ d\theta_2$$
$$= P(\theta_1|Z_1 = z_1^*), \qquad (12)$$

$$P(\theta_2|Z_1,Z_2,X,\theta_3) \approx \int_{\theta_1} P(\theta_1|Z_1,Z_2,X,\theta_3) \cdot P(\theta_2|Z_2,X,\theta_3)\ d\theta_1$$
$$= P(\theta_2|Z_2 = z_2^*). \qquad (13)$$

$z_1^*$ and $z_2^*$ make $\theta_1$ and $\theta_2$ conditionally independent. As the dependency structure of $\theta_1$, $Z_1$ and $Z_2$ is head-to-tail, the relationships $\theta_1 \perp\!\!\!\perp Z_2|Z_1$ and $\theta_1 \perp\!\!\!\perp X|Z_1$ hold, which is known as d-separation (Geiger et al., 1990). In the same manner, the relationships $\theta_2 \perp\!\!\!\perp Z_1|Z_2$ and $\theta_2 \perp\!\!\!\perp X|Z_2$ hold. Eq. (12) and (13) indicates that, where $z_1^*$ and $z_2^*$ are available, the posterior distributions of model parameters (or likelihood of model parameters) are not dependent on $X_{obs}$, other parameters $\theta$, and other intermediate variables $Z$. Parameter $\theta_1$ depends only on $z_1^*$ and parameter $\theta_2$ only on $z_2^*$. Thus, the quasi-observations can make the parameter estimation drastically easier by removing the complex dependency between random variables while generating the conditional independence. This indicates that we can separately estimate the parameters of each model by using quasi-observations, allowing for typical maximum likelihood estimation (MLE) or Bayesian estimation.

## 4. Demonstration: calibration of an urban freight simulator

Disaggregate data for urban freight operations, such as shipment records, is scarce and challenging to obtain. Data are only collected by using costly and infrequent establishment-based surveys, implemented large scale only for a small number of cities in the past (Hunt et al., 2006; Allen et al., 2012; Alho and e Silva, 2015; Toilier et al., 2016; Cheah et al., 2018; Oka et al., 2019). Therefore, freight modeling typically requires a number of assumptions to compensate the lack of the disaggregated data (de Jong et al., 2016). Given the circumstances, the advancement of calibration methods is important to facilitate the application of ABMs relying on limited survey data. If we can assume that fundamental business practices of establishments (e.g. inventory management and vehicle operations) are similar across cities, having the calibration process relying on screenline counts for a city of interest allows for wider use of an ABM. While the proposed calibration framework is designed to be widely applicable to ABMs for urban transportation in general, this section demonstrates the practicality of SLTC by applying it to SimMobility Freight (Sakai et al., 2020) the urban freight simulator in an agent-based urban transportation simulation platform.

### 4.1. SimMobility and SimMobility Freight

*4.1.1. SimMobility*

Since the framework of SimMobility is described in Adnan et al. (2016), this section provides only a brief overview. The simulations in SimMobility are fully disaggregate and the agents in the simulations are consistent across different temporal scales. In SimMobility, various model components are tightly integrated, which allows for feedback mechanisms which this calibration process leverages.

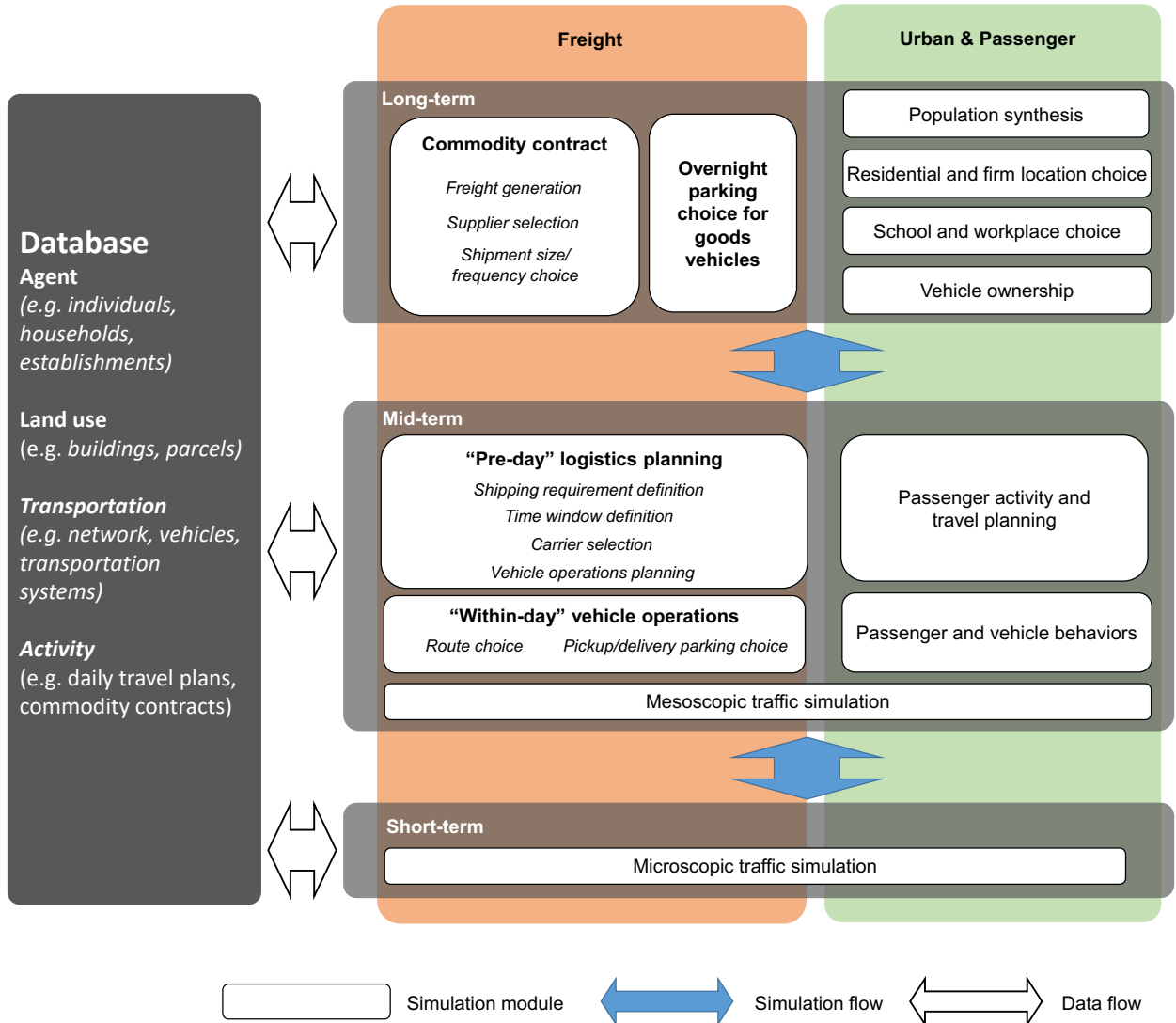

**Fig. 4** *The framework of SimMobility and SimMobility Freight. (Adapted from Sakai et al., 2020)*

As Fig 4 shows, SimMobility consists of three temporal dimensions, namely, Long-Term (LT), Mid-Term (MT) and Short-Term (ST). The LT model focuses on strategic decision making that requires a long-term perspective, such as residential and job locations, establishment locations, vehicle ownership, overnight parking locations, and commodity contracts (i.e. selling and purchasing policies). The MT model covers the decisions that are directly connected to day-level activities, such as activity schedules for passenger travels (Bowman and Ben-Akiva, 2001), logistics planning regarding vehicle operations, route choices, and en-route parking locations. The MT model also incorporates a meso-scale traffic simulator which is computationally efficient and suitable for a large-scale simulation (Lu et al., 2015). Finally, the ST model conducts the micro-simulation of vehicle behaviors on the road network, replicating their behaviors in detail (Azevedo et al., 2017). The ST model represents high spatial temporal resolution (i.e. in the order of tenth of a second) events and decisions, such as lane-changing, braking and accelerating, individual and crowd pedestrian movement. Each of the three levels is modular and autonomous and can be simulated in isolation with appropriate inputs.

*4.1.2. SimMobility Freight*

The group of the components of SimMobility dedicated to urban freight simulations is called SimMobility Freight (Sakai et al., 2020). SimMobility Freight also fits into the temporal framework of SimMobility. The LT model predicts urban commodity flow (through the simulation of commodity contracts) and overnight parking choices for goods vehicles. Since the latter is not the focus of the present paper, "LT model" hereafter simply means the former. The MT model predicts goods vehicle operations.

Commodity flow is predicted through three modules in the LT model: Freight Generation Module (FGM), Supplier Selection Module (SSM), and Size and Frequency Module (SFM).

- FGM predicts the annual production and consumption (the amounts to ship and receive) for each establishment.

- SSM translates the annual quantity of consumption into contract-based demand and matches each contract-based demand with a supplier; a supplier selection model predicts that selection of a supplier for a receiver at the contract level.

- FM predicts the shipment size and frequency for each contract.

The main input to the above three modules is the list of establishments located in the study area, with the details of employment, floor area, industry type, and function type and the output is the commodity contracts, defining daily shipments with the information of origin and destination, commodity type, and size.

The MT model consists of a series of random utility discrete choice models and heuristics. The MT model, using the list of shipments as the main input, predicts tour-based travel demand, routing, and traffic conditions associated with commodity flow. Figure 5 shows the process of the MT model simulation.

- "Pre-day" logistics planning module consists of four sub-models, Shipment Requirement Definition, Time Window Selection, Carrier Selection, and Vehicle Operation Planning. The set of models predicts vehicle operation plans (VOP) of carriers, specifying vehicle movements in a day, pick-ups, deliveries, and overnight parking, which is, in nature, tour-based travel demand.

- "Within-day" vehicle operations module consists of Route Choice and Pickup/Delivery Parking Choice models, and a mesoscopic traffic simulator. It predicts vehicle routing and traffic conditions.

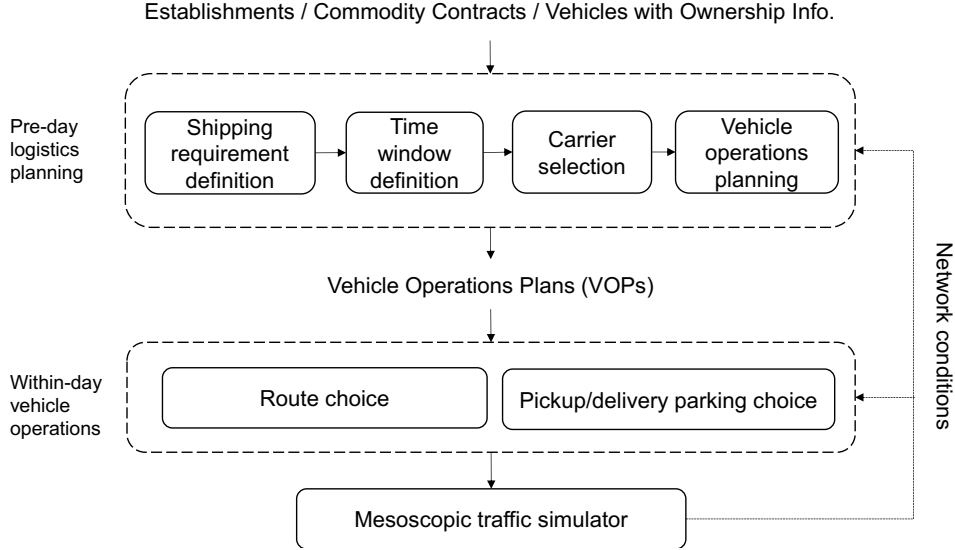

**Fig. 5** *The overview of MT model. (Adapted from Sakai et al., 2020)*

*4.1.3. Models for commodity flow estimation*

Each of FGM, SSM and SFM has a core model which parameters we consider for calibration. We further detail those models, namely Freight Generation model, Supplier Selection model and Shipment Size model.

In Freight Generation model, model parameters $\beta_{prod}$ and $\beta_{cons}$ are set for each establishment group $e$ defined by the types of commodity handled, industry, and function. The following linear functions estimate the production and consumption of establishment $n$ of group $e$:

$$y_{prod}^n = \beta_{const}^{p,e} + \beta_{floor}^{p,e} \cdot x_{floor}^n + \beta_{emp}^{p,e} \cdot x_{emp}^n + \beta_{floor \cdot emp}^{p,e} \cdot x_{floor}^n x_{emp}^n, \qquad (14)$$

$$y_{cons}^n = \beta_{const}^{c,e} + \beta_{floor}^{c,e} \cdot x_{floor}^n + \beta_{emp}^{c,e} \cdot x_{emp}^n + \beta_{floor \cdot emp}^{c,e} \cdot x_{floor}^n x_{emp}^n + \beta_{prod}^{c,e} \cdot y_{prod}^n, \qquad (15)$$

where $x_{floor}^n$ is floor area of establishment $n$, $x_{emp}^n$ is employment size of establishment $n$, parameters $\boldsymbol{\beta_{prod}^e} \equiv \left(\beta_{const}^{p,e}, \beta_{floor}^{p,e}, \beta_{emp}^{p,e}, \beta_{floor \cdot emp}^{p,e}\right)$ and $\boldsymbol{\beta_{cons}^e} \equiv \left(\beta_{const}^{c,e}, \beta_{floor}^{c,e}, \beta_{emp}^{c,e}, \beta_{floor \cdot emp}^{c,e}, \beta_{prod}^{c,e}\right)$ are specific to group $e$.

Supplier Selection model predicts a supplier selection by a receiver for a contract-based demand of a particular commodity type (the detail of the model is available in Sakai et al. (2018)). The probability that a receiver chooses supplier is expressed by error component logit mixture model. The model parameters are set for each combination of the commodity type, receiver function, and supplier function. The utility of supplier $i$ for receiver $n$ is:

$$U_{ni} = V_{ni} + M_{ni} + \varepsilon_{ni}, \qquad (16)$$

where $V_{ni}$ is the systematic component of utility function, $M_{ni}$ is the random component that captures the correlation structure among alternatives, and $\varepsilon_{ni}$ is identically and independently Gumbel distributed random component. $V_{ni}$ and $M_{ni}$ are further specified as follows:

$$V_{ni} = \beta_{time}^{epg} \ln x_{time}^{n,i} + \beta_{prod}^{epg} \ln x_{prod}^i + \beta_{demand}^{epg} \ln x_{demand}^n + \beta_{const}^{epg}, \qquad (17)$$

$$M_{ni} = \begin{cases} \sigma_{or}^{epg}\eta_{or} + \sigma_{dws}^{epg}\eta_{dws}, & \text{if } sfnc_i = office \text{ or } retail \\ \sigma_{lf}^{epg}\eta_{lf} + \sigma_{dws}^{epg}\eta_{dws}, & \text{if } sfnc_i = logistics\ facilities \\ 0, & \text{if } sfnc_i = factory \end{cases} \quad (18)$$

where $x_{time}^{n,i}$ is travel time between receiver $n$ and supplier $i$, $x_{prod}^{i}$ is production of supplier $i$, $x_{demand}^{n}$ is the size of contract-based demand of receiver $n$ in terms of weight, $epg$ is the establishment pair group, defined by commodity type, receiver's and supplier's function types. $\eta_{or}, \eta_{lf}, \eta_{dws}$ are random terms following the standard normal distribution, and $sfnc_i$ indicates the function type of supplier $i$. The model parameter is $\beta_{ssm}^{epg} \equiv (\beta_{time}^{epg}, \beta_{prod}^{epg}, \beta_{demand}^{epg}, \beta_{const}^{epg}, \sigma_{or}^{epg}, \sigma_{lf}^{epg}, \sigma_{dws}^{epg})$.

The random component in the model captures the correlations among potential suppliers. The choice probability of supplier $i$ by receiver $n$ for a contract-based demand is:

$$P_{ni} = \frac{\exp(V_{ni} + M_{ni})}{\sum_{j \in J} \exp(V_{nj} + M_{nj})}. \quad (19)$$

Finally, the Shipment Size model predicts the size of shipment for each contract. The size of shipment of contract $c$, received by establishment $n$ which belongs to group $e$ is:

$$\ln s_{c^n} = \beta_{const}^{e} + \beta_{size}^{e} \ln x_{size}^{c^n} + \beta_{dist}^{e} \ln x_{dist}^{c^n} + \beta_{dense}^{e} \ln x_{dense}^{n}, \quad (20)$$

where $x_{size}^{c^n}$ is a size of contract $c^n$ in terms of commodity weight, $x_{dist}^{c^n}$ is the distance between the supplier and the receiver for contract $c^n$, $x_{dense}^{n}$ is the establishment density at the location of receiver $n$, and $e$ is the group defined by the commodity and receiver's function types. The parameter is $\beta_{shipsize}^{e} \equiv (\beta_{const}^{e}, \beta_{size}^{egp}, \beta_{demand}^{egp}, \beta_{const}^{egp}, \sigma_{or}^{egp}, \sigma_{lf}^{egp}, \sigma_{dws}^{egp})$.

The shipment frequency for contract $c^n$ is defined by the following equation:

$$f_{c^n} = \frac{x_{size}^{c^n}}{s_{c^n}}. \quad \forall c^n \quad (21)$$

The original parameters of the SimMobility Freight LT model were estimated using the data from 2013 Tokyo Metropolitan Freight survey (TMFS), arguably the largest scale establishment survey for urban freight in the past, covering the detail information of establishments (as suppliers, receivers, and carriers) and their shipments.

4.2. Calibration approach

We use SLTC to calibrate the LT model in SimMobility Freight to improve the model fit for policy analyses in Singapore. Specifically, we calibrate model parameters $\beta_{prod}$, $\beta_{cons}$, $\beta_{ssm}$ and $\beta_{shipsize}$ based on the observed full day screenline count data from 2012. Figure 6 shows the simulation flow of SimMobility Freight on the left and the flow of SLTC on the right. A key assumption we made in this application of the concept is that the gap between the observed and simulated traffic is mainly attributable to the prediction of commodity flow and thus the calibration is conducted only for the LT model parameters. To calibrate both the LT and MT (i.e. goods vehicle operation) models, additional data needs to be available. We leave the extension to the simultaneous calibration of the LT and MT models of an urban freight simulator for the future research task.

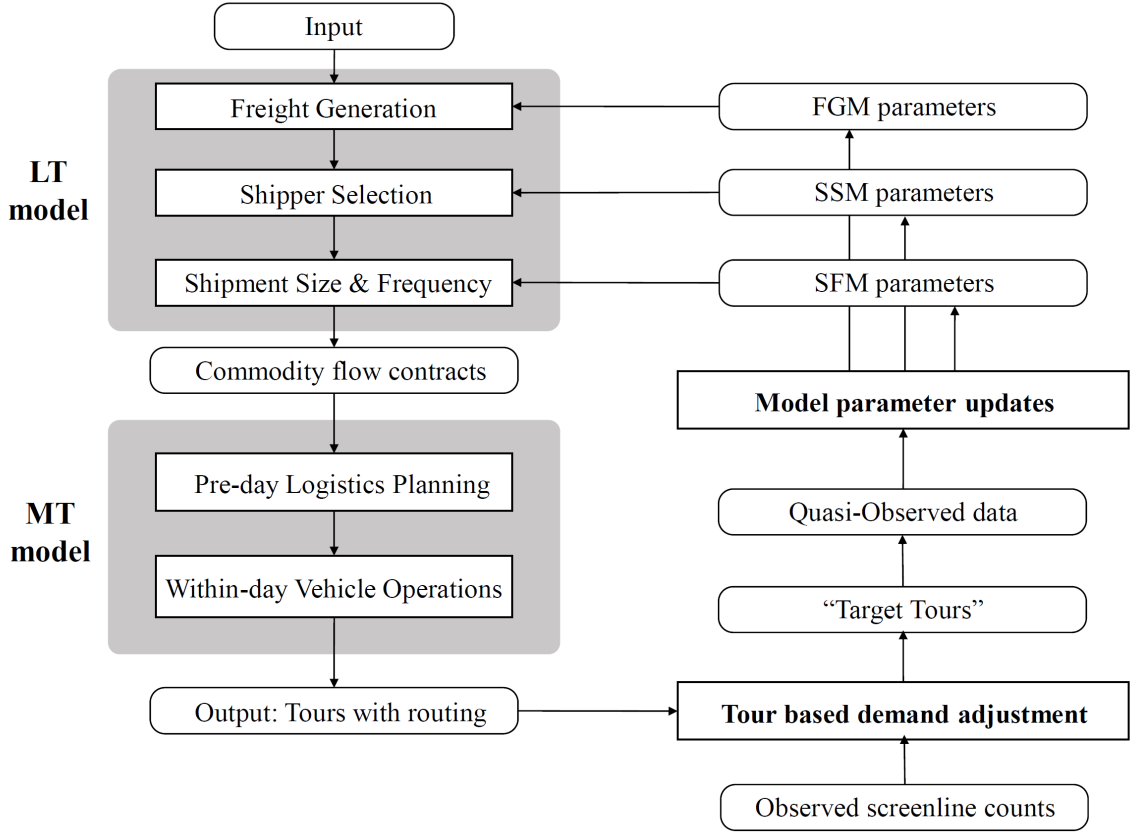

**Fig. 6**  *The overview of screenline-based two-step calibration for SimMobility Freight.*

*4.2.1. Tour-based demand adjustment and penalty parameter identification*

The simulation using initial (or uncalibrated) LT and MT models generates tours with routing information. At first, we make a mapping matrix $A$ based on them. The number of screenlines used for the calibration is 56, the initial number of tours is 199,573, and the initial number of unique SLB tours is 160,952. Therefore, the size of mapping matrix $A$ is 160,952 × 56.

The proposed method requires predefined penalty parameter $\lambda$ to calibrate tours and generate *Target Tours*. It is important to find a proper value for the penalty parameter. While the RMSE decrease as $\lambda$ decreases, $\lambda$ should not be too small as it leads to over-fitting. We use leave-one-out cross-validation (LOOCV) for the analysis to identify the relationship between $\lambda$ and the RMSE, using 55 observed screenline counts for training and the remaining one screenline count for validation. LOOCV is advantageous on minimizing the generalization error and avoiding over-fitting.

*4.2.2. LT model parameter updates with Target Tours*

From *Target Tours*, we generate quasi-observed shipments (QO shipments), which are for updating the LT model parameters. Cloned/removed tours are associated with shipments and contracts. QO shipments are obtained by adjusting the initial shipments based on those associations. Here, it must be noted that sizes of the QO shipments assumed to be the same with those of the initial shipments.

We also obtain quasi-observed contract size (QO contract size) $\hat{x}^{cn}_{size}$ as follows:

$$\hat{x}_{size}^{c^n} = \frac{\hat{f}_{c^n}}{f_{c^n}} \cdot x_{size}^{c^n}, \quad \forall c^n \tag{22}$$

where $f_{c^n}$ and $\hat{f}_{c^n}$ are the frequencies of the initial shipments and the QO shipments, respectively.

We use the QO shipments and/or QO contract sizes to update each of three key sub-models in the LT model as follows:

*Freight Generation model update*

The aggregation of QO contract sizes by supplier and receiver provides quasi-observed production $\hat{y}_{prod}^n$ and consumption $\hat{y}_{cons}^n$; then, the pairs of $(\hat{y}_{prod}^n, x^n)$, $(\hat{y}_{cons}^n, x^n)$ are available for estimating the model parameters of the Freight Generation model, $\beta_{prod}$ and $\beta_{cons}$, by linear regression.

*Supplier Selection model update*

The process of using QO shipments to update the parameters of the Supplier Selection model consists of two steps. First, based on QO shipments, we calculate the probability distribution of shipment origins given a shipment destination as follows:

$$P_{kr} = \frac{q_{kr}}{\sum_{z \in Z} q_{zr}}, \quad \forall k \in Z, \forall r \in Z \tag{23}$$

where $q_{zr}$ indicates the number of QO shipments from zone $z$ to zone $r$.

Next, we reassign (i.e. update) a supplier for each contract. This reassignment process consists of two sub-steps. First, we select the origin zone using the above probability distribution of origins given a receiver's location as the destination. Second, we select a supplier from the selected zone of origin, using the probabilities calculated by Eq. (19). We consider the selected supplier as the quasi-observed supplier for a contract.

The re-estimation of the discrete choice model requires choice sets for each contract. We randomly select alternatives other than the quasi-observed supplier from suppliers that belong to the same establishment pair group $epg$. Each choice set consists of 50 alternatives including selected one (i.e. the quasi-observed supplier). Using this data, we re-estimate the model parameters of mixed logit model in Eq. (17, 18) with the simulated maximum likelihood method.

*Shipment Size model update*

While shipment sizes are not updated from the initial shipment sizes $s_{c^n}$, which is put forward as an assumption in this process, we still update the model parameter of Shipment Size model. For maintaining the shipment sizes from the following iteration similar to those from the initial simulation, the relationship between contract size and shipment size needs to be updated. We use shipment size $s_{c^n}$ and QO contract size $\hat{x}_{size}^{c^n}$ obtained from Eq. (22). As we have the pair of $(\hat{x}_{size}^{c^n}, s_{c^n})$ which includes quasi-contract size $\hat{x}_{size}^{c^n}$ we can estimate the model parameter $\beta_{shipsize}$ by linear regression, similarly to the Freight Generation model.

4.3. Results

*4.3.1 Determination of penalty parameter*

Figure 7 shows the result of LOOCV. The horizontal axis indicates penalty parameter λ (log-scale) and the vertical axis indicates the RMSE from LOOCV. The result shows that, for the penalty parameter of greater than 5000, the smaller the penalty parameter, the lower the RMSE. Such relationship is reversed when the penalty parameters is less than 5000, indicating the situation of over-fitting. As a result, we consider 5000 as the optimal penalty parameter and use it for the following step.

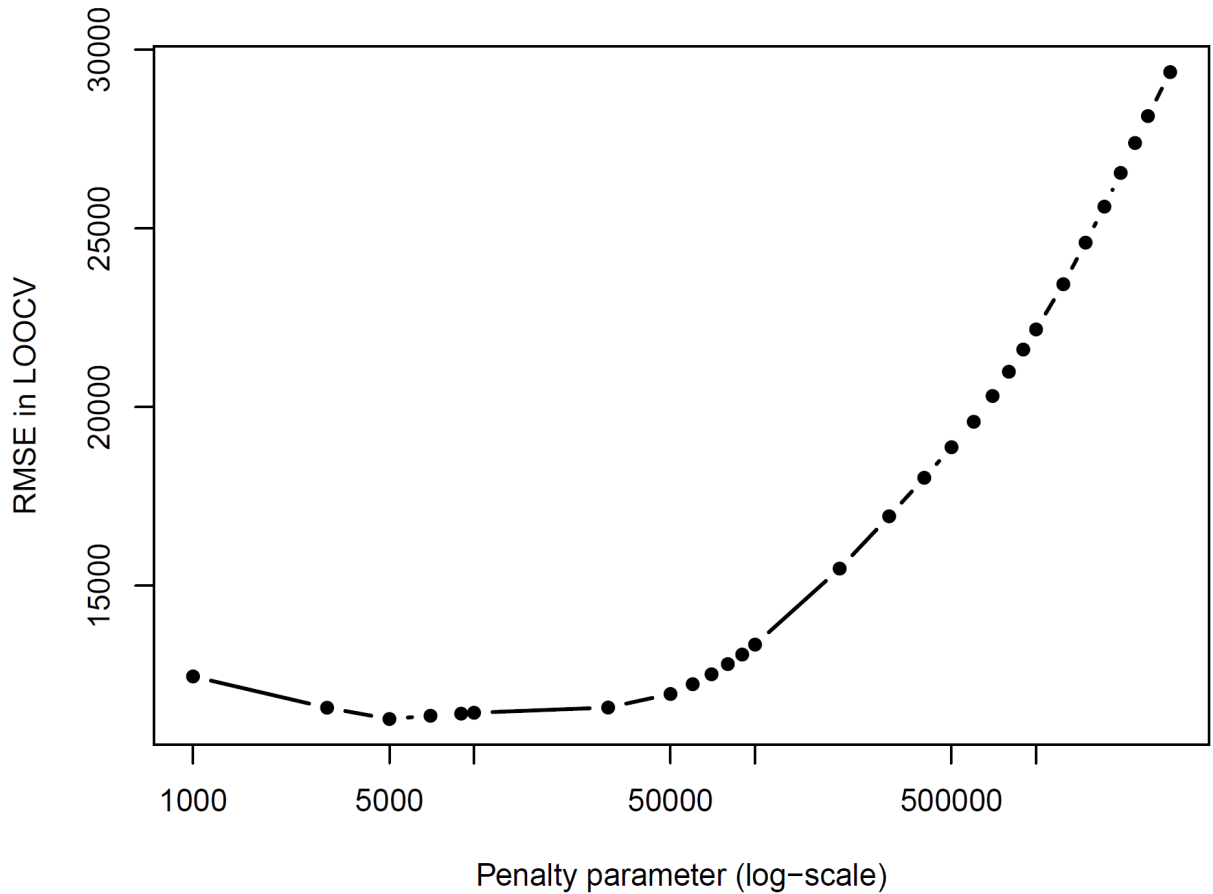

**Fig. 7** *The relationship between penalty parameter and RMSE from LOOCV.*

*4.3.2 Convergence process of SLTC*

Figure 8 shows the flow of SLTC adapted to SimMobility Freight. We take an iterative process and repeat the tour-based demand adjustment and the model parameter updates until convergence. For $k = 1$, we only execute the right side of the figure and use the initial model parameters which are estimated based on the 2013 TMFS data; we simply run Freight Generation model (production model and consumption model), Supplier Selection model, Shipment Size model, and the MT model. The output is the simulated screenline counts of 56 screenlines. From $k = 2$, tour-based demand adjustment generates *Target Tours*, QO shipments, QO contract sizes and the other quasi-observed data, such as quasi-observed production, quasi-observed consumption, and quasi-observed suppliers. Following, the parameter estimation of FGM, SSM and SFM is executed by using these quasi-observed data. To check the convergence, we use the RMSE between simulated screenline counts and observed screenline counts. If the difference of $k$th RMSE and $k-1$th RMSE is less than a predetermined threshold, we terminate the calibration process. Otherwise, we continue the process and increase $k$ by 1.

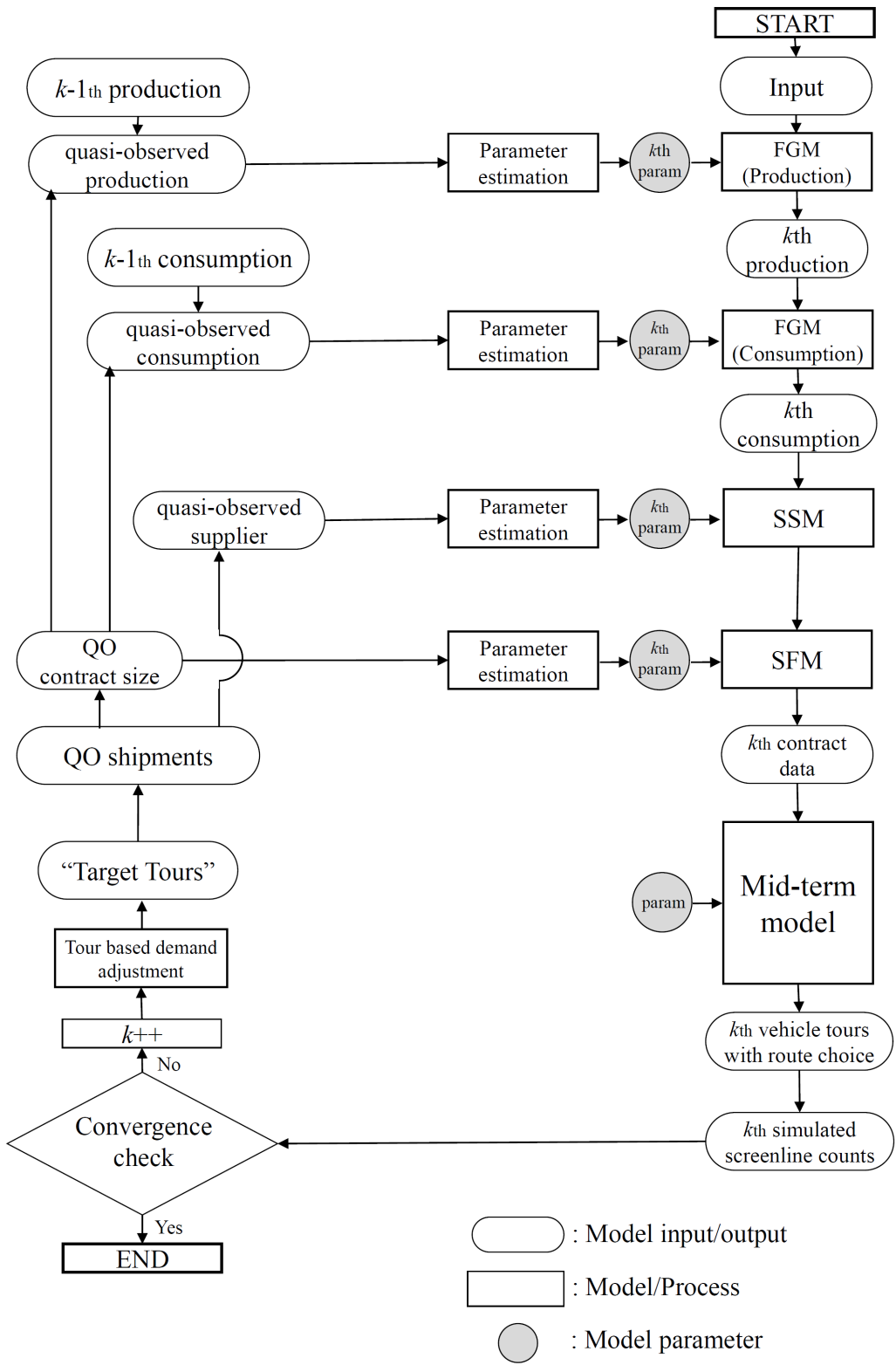

**Fig. 8** The flowchart of two-step calibration method.
"Input"' means the data of establishments and road network. If $k = 1$, $k_{th}$ parameters are the initial parameters and the parameter estimation steps are skipped; otherwise, the whole process is executed.

Figure 9 shows the convergence process of RMSE and mean absolute error (MAE). The initial RMSE and MAE are 32377.93 and 23580.60 respectively. The final RMSE is 14666.78 and final MAE is 9678.14. As the average observed traffic flow of 56 screenlines is 42901.59, the MAE ratio, which is the ratio of MAE to the average observed flow, improved from 0.55 to 0.23 after only 13 iterations.

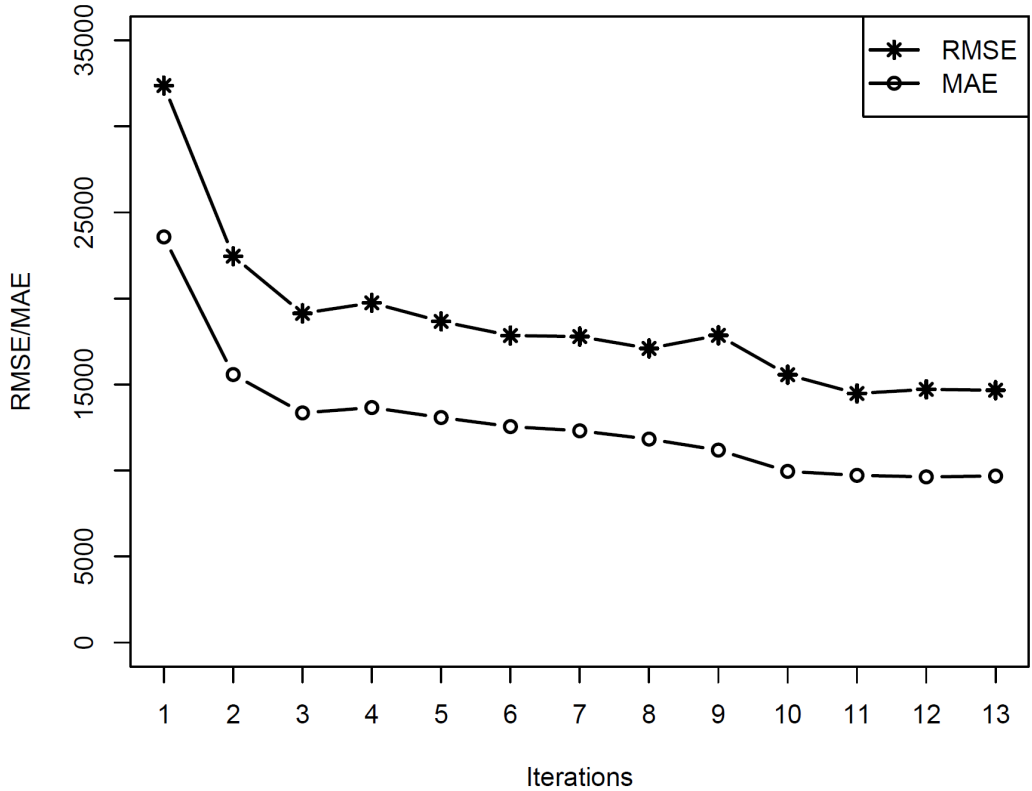

**Fig. 9**   RMSE and MAE variation during convergence process.

*4.3.3 Calibration result*

Figure 10 compares the observed and simulated traffic count of each screenline. The red points show the initial result which uses the original model parameters estimated using the TMFS data. The blue points show the final result which uses the calibrated model parameters by SLTC. Initially, many simulated screenline counts are overestimated; on the other hand, the calibrated parameters improve the predictability significantly. The result underlines that SLTC properly calibrated the model parameters of the LT model in SimMobility Freight.

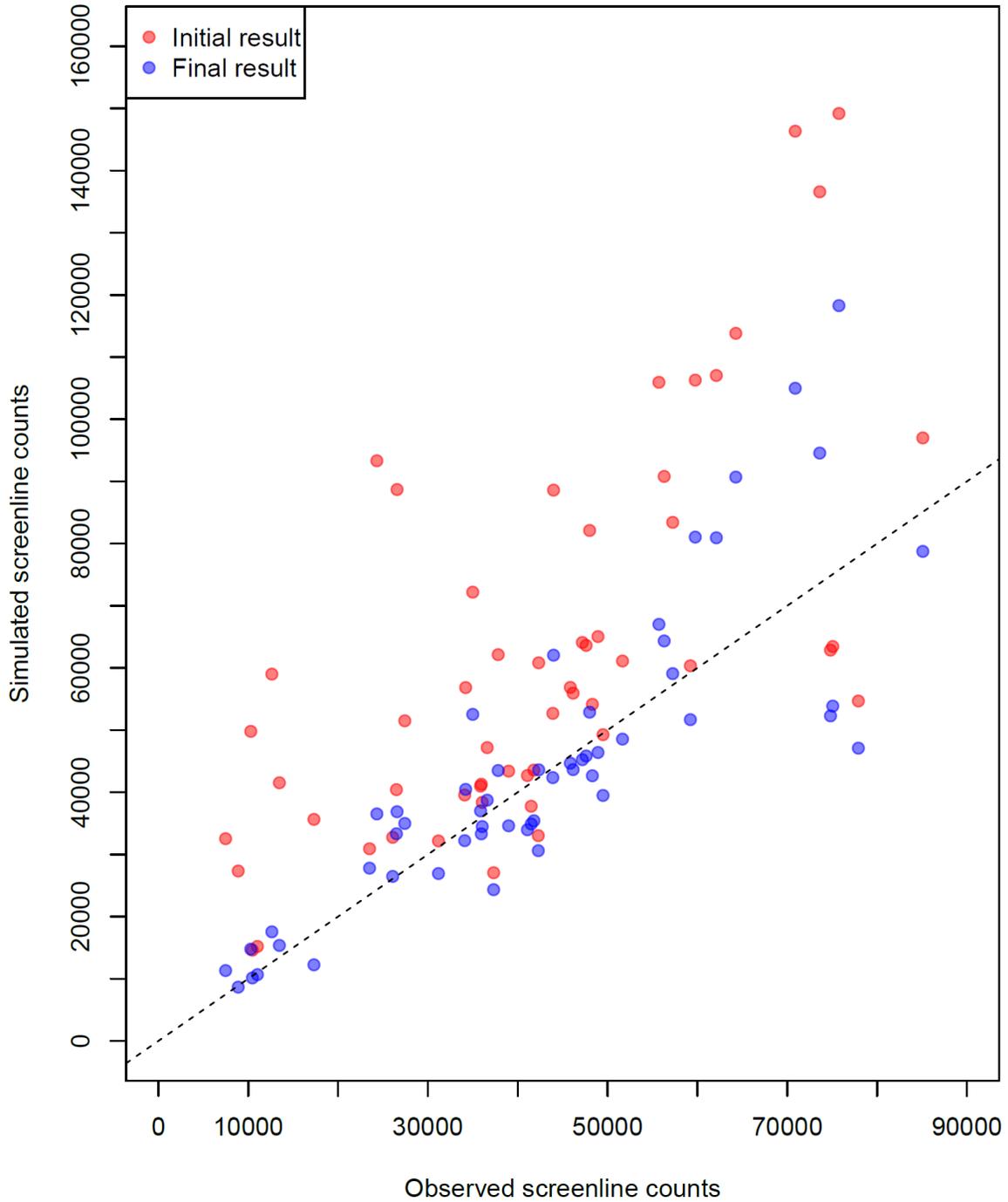

**Fig. 10** Observed and simulated traffic count of each screenline.

## 5. Conclusion

Despite the recent advances in agent-based urban transportation modeling, the systematic approach for calibrating models was lacking. Our claimed contribution is methodological. This research contributes to addressing the shortage of the research focusing on the calibration of those models. SLTC is a powerful method to calibrate an ABM for transportation simulations, especially where the available data for calibration is limited to screenline counts. We discussed the concept of SLB tour and described the methods to generate and use quasi-observed data. For demonstrating the practicality of the method, we used SLTC to calibrate an agent-based urban freight simulator, SimMobility Freight. In this demonstration, freight model parameters were originally estimated using survey data from the Tokyo Metropolitan Area, Japan. These parameters were calibrated and applied to a simulation in Singapore. SLTC allowed us to calibrate the parameters with only 13 iterations using the observed screenline counts in Singapore. Although the transferability of the original models for a study area of interest should be carefully assessed (ideally with supplemental data), the result indicates that SLTC has a potential for promoting the adoption of ABMs as traffic count data has become increasingly available in many cities due to the broader installations of sensing devices. Although it not demonstrated in this paper, the concept of SLTC is also applicable to higher temporal granularity (e.g., screenline counts for each hour) and passenger models.

While this research considers the use of screenline counts, which are widely available for calibration, it is worth developing the method to jointly use the different set of data (e.g., screenline counts and delivery record samples). Regarding to the application in the current research, the MT model parameters are not subject to calibration. The combined use of multiple data sources has a potential to calibrate the commodity flows and vehicle operations, which should enhance the validity of the simulator. As mentioned in Section 4, we leave such extension for the future research.

**Acknowledgements**

This research is supported in part by the Singapore Ministry of National Development and the National Research Foundation, Prime Minister's Office under the Land and Liveability National Innovation Challenge (L2 NIC) Research Programme (L2 NIC Award No L2 NICTDF1-2016-1) and CREATE program, Singapore-MIT Alliance for Research and Technology (SMART) Future Urban Mobility (FM) IRG. Any opinions, findings, and conclusions or recommendations expressed in this material are those of the author(s) and do not reflect the views of the Singapore Ministry of National Development and National Research Foundation, Prime Minister's Office, Singapore. We thank the Urban Redevelopment Authority of Singapore, JTC Corporation and Land Transport Authority of Singapore for their support.